\documentstyle [prl,aps,twocolumn,epsf,psfig]{revtex}
\newskip\humongous \humongous=0pt plus 1000pt minus 1000pt
\def\caja{\mathsurround=0pt}
\def\eqalign#1{\,\vcenter{\openup1\jot \caja
    \ialign{\strut \hfil$\displaystyle{##}$&$
    \displaystyle{{}##}$\hfil\crcr#1\crcr}}\,}
\newif\ifdtup

\relax 

\begin{document}

\newcommand{\newc}{\newcommand}

\newc{\be}{\begin{equation}}
\newc{\ee}{\end{equation}}
\newc{\ba}{\begin{eqnarray}}
\newc{\ea}{\end{eqnarray}}
\newc{\bea}{\begin{eqnarray}}
\newc{\eea}{\end{eqnarray}}
\newc{\D}{\partial}
\newc{\ie}{{\it i.e.} }
\newc{\eg}{{\it e.g.} }
\newc{\etc}{{\it etc.} }
\newc{\etal}{{\it et al.}}

\newc{\ra}{\rightarrow}
\newc{\lra}{\leftrightarrow}
\newc{\no}{Nielsen-Olesen }
\newc{\tp}{'t Hooft-Polyakov }
\newc{\lsim}{\buildrel{<}\over{\sim}}
\newc{\gsim}{\buildrel{>}\over{\sim}}
\draft \preprint{DEMO-HEP-99/01 March 99} 
\title
{Stabilizing Textures in 3+1 Dimensions with Semilocality} 

\bigskip

\author{L. Perivolaropoulos} 

\address{Institute of Nuclear Physics,
N.C.S.R. Demokritos \\ 153 10 Athens; Greece\\ e-mail: 
leandros@mail.demokritos.gr} 

\date{\today}
\maketitle 

\begin{abstract}
\noindent It is shown that textures in 3+1 dimensions can be 
stabilized by partial gauging (semilocality) of the vacuum 
manifold such that topological unwinding by a gauge transformation 
is not possible. This introduction of gauge fields can be used to 
evade Derrick's theorem without higher powers of derivatives and 
prevent the usual collapse of textures by the effective `pressure' 
terms induced by the gauge fields. A virial theorem is derived and 
shown to be a manifestation of the stability of the configuration 
towards collapse. 
\end{abstract}

\narrowtext 

\section{Introduction}

\noindent 

Texture-like topological defects where the topological charge 
emerges by integrating over the whole physical space (not just the 
boundary) have played an important role in both particle physics 
and cosmology. Typical examples are the skyrmion\cite{skyrme} 
which offers a useful effective model for the description of the 
nucleon and the global texture\cite{turok} where an instability 
towards collapse of the scalar field configuration has been used 
to construct an appealing mechanism for structure formation in the 
universe. 

A typical feature of this class of scalar field configurations are 
instabilities towards  field rescalings which usually lead to 
collapse and subsequent decay to the vacuum via a localized highly 
energetic event in space-time. The property of collapse is a 
general feature of global field configurations in 3+1 dimensions 
and was first described by Derrick\cite{derrick}. This feature is 
particularly useful in a cosmological setup because it provides a 
natural decay mechanism which can prevent the dominance of the 
energy density of the universe by texture-like defects. At the 
same time, this decay mechanism leads to a high energy event in 
space-time that can provide the primordial fluctuations for 
structure formation. 

In the particle physics context where a topological defect 
predicted by a theory can only be observed in accelerator 
experiments if it is at least metastable, the above instability is 
an unwanted feature. A usual approach to remedy this feature has 
been to consider effective models where non-renormalizable higher 
powers of scalar field derivatives are put by hand. This has been 
the case in QCD where chiral symmetry breaking is often described 
using the low energy 'pion dynamics' model. Texture-like 
configurations occur here and as Skyrme first pointed out they may 
be identified with the nucleons (Skyrmions)\cite{skyrme}. Here 
textures are stabilized by non-renormalizable higher derivative 
terms in the quantum effective action. However no one has ever 
found such higher derivative terms with the right sign to 
stabilize the Skyrmion. 

An alternative approach to stabilize texture-like configurations 
is the introduction of gauge fields\cite{bt,p99} which can be 
shown to induce pressure terms in the scalar field Lagrangian thus 
balancing the effects of Derrick-type collapse. In the case of 
complete gauging of the vacuum manifold however, it is possible 
for the texture configuration to relax to the vacuum manifold by a 
continuous gauge transformation that can remove all the gradient 
energy (the only source of field energy for textures) from the 
non-singular texture-like configuration\cite{turok}. This 
mechanism of decay via gauge fields is not realized in singular 
defects where the topological charge emerges from the boundaries. 
In these defects, singularities, where the scalar field is 0, can 
not be removed by continuous gauge transformations. 

Recent progress in semilocal defects has indicated that physically 
interesting models can emerge by a partial gauging of the vacuum 
manifold of field theories. This partial gauging (semilocality) 
can lead to new classes of stable defect solutions that can 
persist as metastable configurations in more realistic models 
where the gauging of the vacuum is complete but remains 
non-uniform. A typical example is the semilocal string\cite{va91} 
whose embedding in the standard electroweak model has led to the 
discovery of a class of metastable 2+1 dimensional field 
configurations in this model\cite{vp}. 

In the context of texture-like configurations, the concept of 
semilocality can lead to an interesting mechanism for 
stabilization. In fact the semilocal gauge fields are unable to 
lead to relaxation of the global field gradient energy because 
they can not act on the whole target space. They do however induce 
pressure terms in the Lagrangian that tend to resist the collapse 
induced by the scalar sector. Therefore they have the features 
required for the construction of stable texture-like 
configurations in renormalizable models without the adhoc use of 
higher powers of derivatives. 

The goal of this paper is to demonstrate the stabilization induced 
by semilocal gauge fields in the context of global 
textures\cite{turok} that form during the symmetry breaking 
$O(4)\rightarrow O(3)$. In the semilocal case discussed below the 
$O(3)$ subgroup of the global $O(4)$ symmetry is gauged. 

\section{Virial Theorem} 

\noindent 
Consider first a texture field configuration in 3+1 
dimensions emerging in the context of a field theory describing a 
global symmetry breaking $O(4)\rightarrow O(3)$. This is described 
by a four component scalar field ${\vec Q}=(Q_1,Q_2,Q_3,Q_4)$ 
whose dynamics is described by the potential \be V({\vec 
Q})={\lambda \over 4} ({\vec Q}^2 - F^2)^2 \label{pot31} \ee The 
initial condition ansatz \be {\vec Q}=(\sin\chi \sin\theta 
\sin\varphi ,\sin\chi \sin\theta \cos\varphi,\sin\chi 
\cos\theta,\cos\chi) \ee with $\chi (r)$ varying between 0 and 
$\pi$ as $r$ goes from 0 to infinity and $\theta$, $\varphi$ 
spherical polar coordinates, describes a configuration that winds 
once around the vacuum $M_0=S^3$ as the physical space is covered. 
Since $\pi_3(M_0) \neq 1$ this is a nontrivial configuration which 
is topologically distinct from the vacuum.  The energy of this 
configuration is of the form 
\be
E=\int_{-\infty}^{+\infty}{1\over 2} ({\vec \nabla}{\vec Q})^2 + 
V({\vec Q})\equiv T + V \ee where we have allowed for possible 
small potential energy excitations during time evolution. A 
rescaling of the spatial coordinate $r \rightarrow \alpha r$ of 
the field ${\vec Q}(r)$ leads to $E_\alpha = \alpha^{-1} T + 
\alpha^{-3} V$ which is monotonic with $\alpha$ and leads to 
collapse, highly localized energy and eventual unwinding of the 
configuration. These highly energetic and localized {\it events} 
in spacetime have provided a physically motivated mechanism for 
the generation of primordial fluctuations that gave rise to 
structure in the universe\cite{turok}. 

The possible stabilization of these collapsing configurations 
could lead to a cosmological overabundance and a cosmological 
problem similar to the one of monopoles, requiring inflation to be 
resolved. At the same time however it could lead to observational 
effects in particle physics laboratories. There are at least two 
ways to stabilize a collapsing texture in 3+1 dimensions. The 
first is well known and includes the introduction of higher powers 
of derivative terms in the energy functional. These terms scale 
like $\alpha^p$  ($p>0$) with a rescaling and can make the energy 
minimization possible thus leading to stable {\it skyrmions}. 
Stable {\it Hopfions}\cite{Hopfions} (solitons with non-zero Hopf 
topological charge) have also been constructed recently by the 
same method. 

The second method of stabilization is less known (but see ref 
\cite{bt,p99} for applications in 2+1 dimensions) and can be 
achieved by introducing gauge fields that partially cover the 
vacuum manifold. In particular, the simplest Lagrangian that 
accepts semilocal texture configurations in 3+1 dimensions can be 
written as follows: 
 
\begin{equation}
{\cal L} = -{1\over 4} G^a_{i j}G^a_{i j} - {1\over 2} D_i Q_a D_i 
Q_a - {1\over 2} D_i Q_4 D_i Q_4 - V \label{model2} 
\end{equation}
where ($i,j=1,2,3$), 
\be
V={1\over 2} \mu^2 (Q_a^2 +Q_4^2)+{1\over 8} \lambda (Q_a^2 
+Q_4^2)^2 \label{ptl3d} \ee and ${\vec Q}\equiv (Q_a,Q_4)$ is a 
four component scalar field ($a=1,2,3$) with vacuum ${\vec Q}^2 = 
F^2$ with $F^2\equiv -{{2 \mu^2}\over \lambda}$. Here we are using 
the notation of Ref. \cite{th74} \ie \be G^a_{i j}\equiv 
\partial_i W^a_j - \partial_j W^a_i + e \epsilon_{abc} W^b_i W^c_j
\ee and \be D_i Q_a=\partial_i Q_a + e \epsilon_{abc} W^b_i Q_c  
\ee 

The field ansatz that describes a semilocal texture may be written 
as 
\begin{equation}
\eqalign{ 
 Q_a &= r_a Q(r) \cr
 Q_4 &= P(r) \cr
 W^a_i &= \epsilon_{i a b} r_b W(r)
\label{texture-ansatz} } 
\end{equation}

The asymptotic behavior of the field functions $Q(r)$, $W(r)$ and 
$P(r)$ can be obtained by demanding finite energy, 
single-valueness of the fields and non-trivial topology. Thus we 
obtain the following asymptotics 
\begin{equation}
\eqalign{ r Q(r) &\rightarrow 0 \cr -e r^2 W(r) &\rightarrow 0 \cr 
P (r) &\rightarrow F   \label{asympinf} } 
\end{equation}
for $r\rightarrow \infty$ and 
\begin{equation}
\eqalign{ r Q(r) &\rightarrow 0 \cr -e r^2 W(r) &\rightarrow 0 \cr 
P (r) &\rightarrow -F   \label{asymp0} } 
\end{equation} 
for $r\rightarrow 0$. We now rescale the coordinate $r$ as 
$r\rightarrow r/eF$ and define the rescaled  fields 
\begin{equation}
\eqalign{q(r) &\equiv r Q(r) F \cr p(r) &\equiv P(r) F \cr 
w(r)&\equiv -W(r)/e r^2 \label{rescfileds} } 
\end{equation}
Using these rescalings, the energy of the semilocal texture ansatz 
(\ref{texture-ansatz}) may be written as \be 4\pi e^{-1} F 
\int_0^\infty dr \; {\cal E}\ee where the energy density ${\cal 
E}$ may be expressed in terms of the rescaled fields as 
\begin{equation}
\eqalign{{\cal E} &= w'^2 + {{2 w^2}\over r^2}(1 - w + {w^4 \over 
4}) \cr &+ {1\over 2}r^2 (q'^2 + p'^2) + q^2 (1-w)^2 + {1\over 8} 
r^2 \beta (q^2 + p^2 -1)^2   \label{endensity} } 
\end{equation}
where $\beta \equiv {\lambda \over e^2}$. By varying the energy 
density with respect to the field functions $w$, $q$ and $p$ it is 
straightforward to obtain the field equations for these functions. 
These may be written as 
\begin{equation}
\eqalign{w'' - {{2 w} \over r^2} + {{3w^2}\over r^2} - {w^3 \over 
r^2} + q^2 (1-w) &= 0 \cr  (r^2 q')' - 2 q (1-w)^2 - {\beta \over 
2} r^2 q (p^2 +q^2 -1) &=0 \cr (r^2 p')' - {\beta \over 2} r^2 p 
(p^2 +q^2 -1) &=0  \label{fieldeqs}} 
\end{equation} 
with boundary conditions 
\begin{equation}
\eqalign{r&\rightarrow 0: \; \; w\rightarrow 0, \; \;  
q\rightarrow 0, \; \; p\rightarrow 1 \cr r&\rightarrow \infty: \; 
\; w\rightarrow 0, \; \;  q\rightarrow 0, \; \; p\rightarrow -1 
\label{boundcond}} 
\end{equation}
The stability towards collapse of the semilocal texture that 
emerges as a solution of the system (\ref{fieldeqs}) with the 
boundary conditions (\ref{boundcond}) can be studied by examining 
the behavior of the energy after a rescaling of the spatial 
coordinate $r$ to $r\rightarrow \alpha r$. For global textures, 
this rescaling leads to a monotonic increase of the energy with 
$\alpha$ (Derrick's theorem) indicating instability towards 
collapse. It will be shown that this instability is not present in 
the field configuration of the semilocal texture due to the 
outward pressure induced by the gauge field which has the same 
effect as the higher powers of derivatives present in the 
skyrmion. 

The rescaling $r\rightarrow \alpha r$ modifies the total energy as 
follows \be E \rightarrow \alpha E_1  + \alpha^{-1}  E_2    + 
\alpha^{-3} E_3 \label{rescenergy} \ee where   
\begin{equation}
\eqalign{E_1 &=A \int_0^\infty  dr \; [w'^2 + {{2 w^2}\over r^2}(1 
- w + {w^4 \over 4})] \cr E_2 &=A \int_0^\infty  dr \;  [{1\over 
2}r^2 (q'^2 + p'^2) + q^2 (1-w)^2] \cr    E_3 &=A \int_0^\infty  
dr \; [ {1\over 8} r^2 \beta (q^2 + p^2 -1)^2]   \label{enterms} } 
\end{equation}
and $A \equiv 4\pi e^{-1} F$. The expression (\ref{rescenergy}) 
has an extremum with respect to $\alpha$ which is found by 
demanding ${{\delta E} \over {\delta \alpha}}|_{\alpha = 1} = 0$. 
This leads to a virial theorem connecting the energy terms $E_1$, 
$E_2$ and $E_3$ as \be E_1 = E_2 + 3 E_3 \label{virial}\ee By 
considering the second variation of the energy with respect to the 
rescaling parameter $\alpha$ it is easy to see that the extremum 
of the energy corresponding to the semilocal texture solution is 
indeed a minimum. Indeed \be {{\delta^2 E} \over {\delta 
\alpha}^2}|_{\alpha = 1} = 2 E_2 + 12 E_3 > 0 \ee 

\section{Conclusion} 

\noindent 
We therefore conclude that the semilocal texture field 
configuration is a local minimum of the energy functional with 
respect to coordinate rescalings in contrast to its global 
counterpart. In addition it is impossible to unwind this 
configuration to the vacuum by a gauge transformation because only 
the $O(3)$ sub-group of the full symmetry $O(4)$ is gauged and 
therefore it is impossible to 'rotate' all four components of the 
scalar field ${\vec Q}$ by a gauge transformation. Therefore an 
initial field configuration with non-trivial $\pi_3(S^3)$ topology 
can neither unwind continously to the vacuum (due to non-trivial 
topology and insufficient gauge freedom) nor collapse (due to the 
derived virial theorem). Thus the energy of the configuration will 
remain trapped and localized either in the form of a static 
configuration (if a solution to the static system (\ref{fieldeqs}) 
exists) or in the form of a localized time-dependent breather-type 
configuration.  In both cases there can be interesting 
consequences for both particle physics and cosmology. 

Interesting extensions of this brief report include the study of 
the cosmological effects of semilocal textures and the possibility 
of embedding these objects in realistic extensions of the standard 
model. It is also important to perform a detailed numerical 
construction of these objects.  A detailed study including these 
and other issues is currently in progress. 

\section{Acknowledgements}
I thank T. Tomaras for useful discussions.

\vfill 
\eject 
\end{document}